%% file: main.tex
\documentclass[10pt]{article}
\usepackage{colm2024_conference}
\input{macro}

\input{semloc-theme}

\title{\tool: Turning Free-Form LLM Reasoning into Executable and Verifiable Semantic Constraints}
\author{%
\textbf{Zhaorui Yang}$^{1,*}$\quad
\textbf{Haichao Zhu}$^{2}$\quad
\textbf{Qian Zhang}$^{1}$\\[3pt]
\textbf{Rajiv Gupta}$^{1}$\quad
\textbf{Ashish Kundu}$^{3}$\\[7pt]
{\small $^{1}$University of California, Riverside\quad $^{2}$Qualcomm\quad $^{3}$Cisco Research}\\[5pt]
}
\date{}

\begin{document}

\maketitle

\begingroup
\renewcommand{\thefootnote}{*}
\footnotetext{Corresponding author:
  \href{mailto:zyang247@ucr.edu}{zyang247@ucr.edu}.}
\endgroup

\begin{abstractbox}\begin{abstract}\input{sections/abstract}\end{abstract}\end{abstractbox}

\begin{figure*}[h]
    \includegraphics[width=\textwidth]{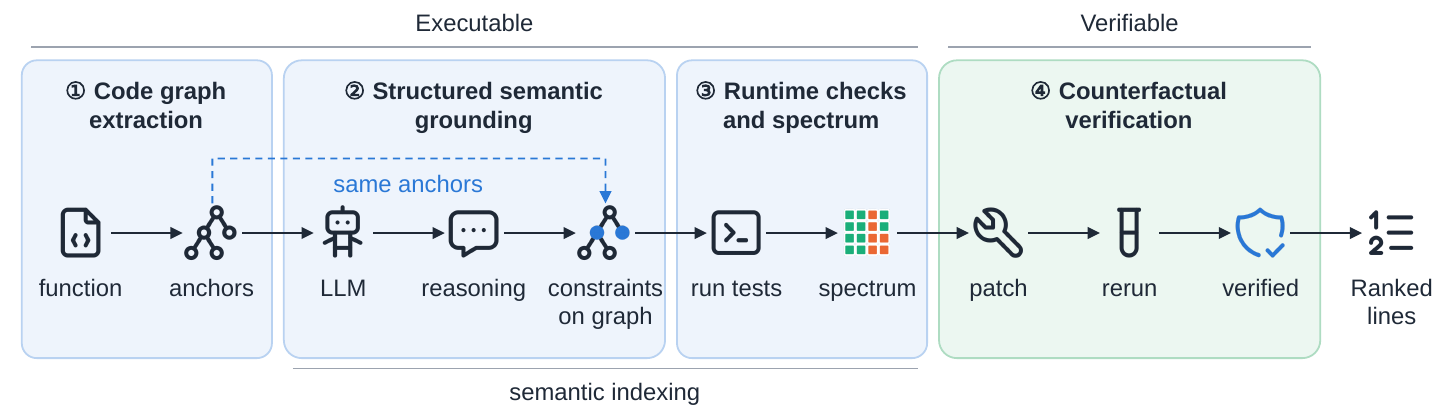}
    \caption{Overview of \tool.}
    \label{fig:design}
\end{figure*}

\section{Introduction}\label{sec:intro}
\input{sections/introduction}

\section{Motivating Example}\label{sec:motivation}
\input{sections/motivating}

\section{Approach}\label{sec:approach}
\input{sections/approach}

\section{Evaluation}\label{sec:evaluation}
\input{sections/evaluation}

\section{Related Work}\label{sec:related}
\input{sections/related_work}

\section{Conclusion}\label{sec:conclusion}
\input{sections/conclusion}

\section*{Data Availability}
The package at \url{https://doi.org/10.5281/zenodo.23119381} contains \sembugs (with each program's repository, commit, and license), \tool, all run outputs, and the scripts for all tables and figures; the models are \texttt{gpt-5.5}, \texttt{openai/\allowbreak gpt-oss-120b}, \texttt{Qwen/\allowbreak Qwen3.8-27B}, and \texttt{Qwen/\allowbreak Qwen3.8-\allowbreak Flash-Next-FP8} (September 2026).

\clearpage
\bibliographystyle{ACM-Reference-Format}
\bibliography{fse26ivr}

\end{document}

%% file: macro.tex
\usepackage[frozencache=true,cachedir=minted-cache]{minted}

\usepackage{microtype}
\usepackage{xspace}
\usepackage{tikz}
\usetikzlibrary{pgfplots.groupplots}

\usepackage{pifont}
\usepackage{booktabs}
\usepackage{pgfplots}
\usepackage{pgfplotstable}
\usepackage{amsmath,amsfonts,amsthm,enumitem}
\usepackage{algpseudocode}
\usepackage[linesnumbered,ruled]{algorithm2e}
\usepackage{graphicx}
\usepackage{url}

\usepackage{textcomp}
\usepackage{xcolor}
\usepackage[labelfont=bf]{caption}
\usepackage{subcaption}
\usepackage[T1]{fontenc}
\usepackage[utf8]{inputenc}

\usepackage[linesnumbered,ruled]{algorithm2e}
\usepackage{colortbl}
\usepackage{tabularx}
\definecolor{semlocrow}{HTML}{E4EEFA}
\usepackage{tcolorbox}
\usepackage{listings}
\usepackage{makecell}

\usetikzlibrary{patterns,arrows}
\usepackage{flushend}
\usepackage{balance}
\usepackage{multicol}
\usepackage{multirow}

\usetikzlibrary{trees}

\usepackage[]{mdframed}
\mdfdefinestyle{finding}{%
  linecolor=black!40,
  linewidth=0.5pt,
  backgroundcolor=gray!8,
  innertopmargin=4pt,
  innerbottommargin=4pt,
  innerleftmargin=6pt,
  innerrightmargin=6pt,
  skipabove=6pt,
  skipbelow=4pt,
}

\usepackage{pgfplots}
\usetikzlibrary{patterns, pgfplots.groupplots}

\usepackage{xspace}
\usepackage{tkz-kiviat} 
\usetikzlibrary{arrows}

\usepackage{adjustbox}

\definecolor{dkgreen}{rgb}{0,0.6,0}
\definecolor{gray}{rgb}{0.5,0.5,0.5}
\definecolor{mauve}{rgb}{0.58,0,0.82}
\definecolor{orangep}{rgb}{0.71, 0.43, 0.89}
\definecolor{orp}{rgb}{1, 0.7, 0.278}

\definecolor{darkBlue}{rgb}{0.000000,0.000000,0.545098}
\definecolor{darkGreen}{rgb}{0.000000,0.392157,0.000000}
\definecolor{DarkGray}{gray}{0.4}
\definecolor{javared}{rgb}{0.6,0,0} % for strings
\definecolor{javagreen}{rgb}{0.25,0.5,0.35} % comments
\definecolor{javapurple}{rgb}{0.5,0,0.35} % keywords
\definecolor{javadocblue}{rgb}{0.25,0.35,0.75} % javadoc
\definecolor{lightgray}{gray}{0.95}
\definecolor{shadecolor}{RGB}{150,150,150}
\definecolor{blueA}{RGB}{204,229,255}
\definecolor{redA}{RGB}{112,0, 0}
\lstdefinestyle{MyCSmallStyle} {
  language=C++,
  frame=none,
  xleftmargin=15pt,
  stepnumber=1, 
  numbers=left, 
  numbersep=5pt,
  numberstyle=\tiny\color[black]{0.177}, 
  belowcaptionskip=\bigskipamount,
  captionpos=b, 
  escapeinside={*'}{'*},
  tabsize=5,
  emphstyle={\bf},
  escapechar=!,
  basicstyle=\scriptsize\ttfamily,
  keywordstyle=\color{javapurple}\bfseries,
  stringstyle=\color{javared},
  commentstyle=\color{javagreen},
  morecomment=[s][\color{javadocblue}]{/**}{*/},
  showspaces=false,
  columns=flexible,
  showstringspaces=false,
  morecomment=[l]{//},
  tabsize=2,
  breaklines=true,
  moredelim=[is][\underbar]{^}{^}
}

\lstdefinestyle{MyJavaSmallStyle} {
  language=Java,
  frame=none,
  xleftmargin=15pt,
  stepnumber=1, 
  numbers=left, 
  numbersep=5pt,
  numberstyle=\color{DarkGray}, 
  belowcaptionskip=\bigskipamount,
  captionpos=b, 
  escapeinside={*'}{'*},
  tabsize=5,
  emphstyle={\bf},
  basicstyle=\scriptsize\ttfamily,
  keywordstyle=\color{javapurple}\bfseries,
  stringstyle=\color{javared},
  commentstyle=\color{javagreen},
  morecomment=[s][\color{javadocblue}]{/**}{*/},
  showspaces=false,
  columns=flexible,
  showstringspaces=false,
  morecomment=[l]{//},
  tabsize=2,
  breaklines=true,
  moredelim=[is][\underbar]{^}{^}
}

\lstdefinelanguage{Scala}{
  keywords={typeof, new, true, false, catch,def,val, function, return, null, catch, switch, var, if, in, while, do, else, case, break, assert, static, void ,declare, const, for, define,fun, ite,class, not, check,sat,String, Int, ArrayList},
  keywordstyle=\color{blue}\bfseries,
  ndkeywords={ export,extends, boolean, throw, implements, import, this, abstract,reduceByKey, reduce, filter, map, reduceByKey, join, Join1, public },
  ndkeywordstyle=\color{mauve}\bfseries,
  otherkeywords={+, =>,<=, ==, >,< , ||},
  identifierstyle=\color{black},
  sensitive=false,
  comment=[l]{//},
  morecomment=[s]{/*}{*/},
  commentstyle=\color{purple}\ttfamily,
  stringstyle=\color{red}\ttfamily,
  morestring=[b]',
  morestring=[b]"
}

\newcommand{\MyPara}[1]{\vspace{.3em}\noindent\textbf{\textit{#1}}~~}

\newcommand{\codefontsmall}[1]{{\small\texttt{#1}}}

\newcommand{\tool}{{{\textsc{SemLoc}}}\xspace}

\newcommand{\toolnocf}{\tool w/o CF\xspace}
\newcommand{\toolnoidx}{\tool w/o semantic indexing\xspace}
\newcommand{\toolcand}{\tool candidate lines\xspace}
\newcommand{\directllm}{Direct LLM\xspace}
\newcommand{\sembugs}{SemBugs\xspace}
\newcommand{\pctflag}{\%Flagged\xspace}
\DeclareUnicodeCharacter{1EA5}{\'{\^{a}}}
\DeclareUnicodeCharacter{1EC5}{\~{\^{e}}}

\newcommand{\rqanswer}[2]{\begin{mdframed}[linewidth=0.5pt,innerleftmargin=6pt,innerrightmargin=6pt,innertopmargin=4pt,innerbottommargin=4pt,skipabove=6pt,skipbelow=4pt]\noindent\textit{Answer to RQ#1.} #2\end{mdframed}}
\newcolumntype{R}[1]{>{\raggedleft\arraybackslash}p{#1}}

%% file: semloc-theme.tex
\definecolor{SemNavy}{HTML}{213C55}
\definecolor{SemTeal}{HTML}{187C80}
\definecolor{SemMuted}{HTML}{637786}
\definecolor{SemPale}{HTML}{F0F6F7}
\definecolor{SemLine}{HTML}{CDDDE2}
\definecolor{semlocrow}{HTML}{EAF3F4}
\hypersetup{colorlinks=true,linkcolor=SemNavy,citecolor=SemTeal,urlcolor=SemTeal,
 pdfauthor={Zhaorui Yang, Haichao Zhu, Qian Zhang, Rajiv Gupta, Ashish Kundu},
 pdftitle={SemLoc: Turning Free-Form LLM Reasoning into Executable and Verifiable Semantic Constraints}}
\setcitestyle{numbers,square,comma,sort&compress}
\tcbuselibrary{most}
\newtcolorbox{abstractbox}{enhanced,breakable,colback=SemPale,colframe=SemLine,
 boxrule=0.4pt,arc=1.5mm,left=10pt,right=10pt,top=5pt,bottom=5pt,
 borderline west={2pt}{0pt}{SemTeal},before skip=8pt,after skip=12pt}
\renewenvironment{abstract}{\noindent{\color{SemNavy}\bfseries Abstract}\par\small}{\par}
\renewcommand{\rqanswer}[2]{\begin{mdframed}[backgroundcolor=SemPale,linecolor=SemTeal,
 linewidth=0.8pt,topline=false,rightline=false,bottomline=false,
 innerleftmargin=9pt,innerrightmargin=9pt,innertopmargin=6pt,innerbottommargin=6pt,
 skipabove=8pt,skipbelow=6pt]\textbf{\color{SemTeal}Answer to RQ#1.} #2\end{mdframed}}
\setminted{rulecolor=\color{SemLine}}
\makeatletter
\renewcommand{\headrulewidth}{0.5pt}
\renewcommand{\headrule}{{\color{SemLine}\hrule\@height\headrulewidth\@width\headwidth\vskip-\headrulewidth}}
\fancypagestyle{firstpage}{\fancyhf{}
 \fancyhead[L]{\includegraphics[height=23pt]{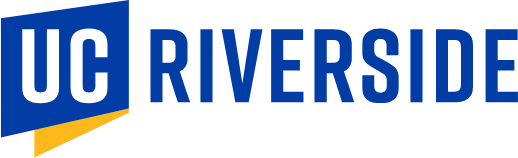}\hspace{12pt}\includegraphics[height=23pt]{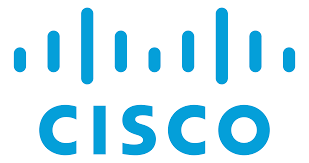}}
 \fancyhead[R]{\small\color{SemMuted} \today}
 \fancyfoot[C]{\small\color{SemMuted}\thepage}}
\renewcommand{\@maketitle}{\vspace*{7pt}\begingroup\centering
 {\fontsize{19}{23}\selectfont\bfseries\color{SemNavy}\@title\par}
 \vspace{12pt}{\normalsize\@author\par}\endgroup\vspace{5pt}\thispagestyle{firstpage}}
\makeatother

\pgfplotsset{compat=1.18}

%% file: sections/abstract.tex
Large language models (LLMs) can describe what a piece of code should compute, but they do so in free-form text.
Such text cannot be checked against program executions, and it does not reveal whether a statement is correct.
We present \tool, which turns LLM reasoning into executable, verifiable semantic constraints.
With structured semantic grounding, the LLM writes each property as a constraint with a fixed category, attached to an anchor in the code graph, such as a variable definition, branch, loop, or return.
\tool inserts each constraint as a runtime check and runs the tests, which yields a semantic violation spectrum over passing and failing tests.
Counterfactual verification then patches the code at an anchor and reruns the tests; a patch that a constraint suggests counts as verified only if all tests then pass.

We study \tool on fault localization, where the tests serve as the oracle for verification, with four LLMs on QuixBugs and on \sembugs, a new dataset of 809 real bugs from 750 GitHub projects.
The constraints are executable: 99.4\% become runtime checks, and on programs where coverage ties the faulty line with at least half of the covered lines, their violations alone rank it first in 40.9\% of the cases, compared with 2.2\% for spectrum-based fault localization and 7.0\% for Daikon invariants.
They make the LLM's answers verifiable: 68.4\% of \tool's top-ranked answers carry a patch that the tests confirm, and 93.6\% of these are correct, which the LLM's own confidence cannot match.
\tool's first-rank accuracy does not differ significantly, for any of the four LLMs, from that of an LLM given the same inputs (81.4\% vs.\ 82.0\%), and it far exceeds that of an LLM that sees only the function (50.0\%).

%% file: sections/introduction.tex
To localize a fault, a developer compares what a program does with what it is supposed to do.
This is one of the most tedious and time-consuming activities in debugging~\cite{fl}.
Most automated techniques rely on signals derived from the structure of executions, which we call \emph{syntactic spectra}.
Spectrum-based fault localization (SBFL) correlates statement coverage with test outcomes~\cite{tarantula,OchiaiSbflCov}, slicing follows data and control dependences~\cite{programSlicing,dynamicSlicing}, and delta debugging isolates the difference between a passing and a failing run~\cite{zellerDD}.
All of them assume that a failing run executes the program differently from the passing runs.

\MyPara{Semantic faults.}
Many bugs do not meet this assumption.
When the fault is a wrong operator, operand, or constant inside an expression, passing and failing tests often execute the same statements in the same order, and only the computed values differ.
The softmax function in Listing~\ref{lst:softmax} is an example: it adds the maximum to each input instead of subtracting it, and every test runs the same lines.
Such faults are common; TSSB-3M alone contains more than three million such fixes from open-source Python projects~\cite{tssb}.
In the 809 real bugs of our dataset, SBFL fails to rank the faulty line first in 83.8\% of the programs even with its best formula.
In 185 programs (22.9\%), the faulty line is tied with at least half of all covered lines. On this particularly ambiguous subset, Ochiai, a widely used SBFL formula, fails to rank the faulty line first in 97.8\% of cases.
For these bugs, the signal lies less in which code executes than in the values it computes.

\MyPara{LLM reasoning.}
Large language models (LLMs) can describe these semantic properties.
For the softmax function in Listing~\ref{lst:softmax}, an LLM can explain that the shifted values should not be positive and that the output should be a probability distribution.
LLM-based fault localization uses this ability to predict suspicious locations, sometimes with an explanation in natural language~\cite{autofl,agentfl,flexfl}.
The reasoning behind the prediction, however, is neither checked against the program's behavior nor compared across passing and failing tests.
A recent survey of code-based agents lists semantic verification beyond executable feedback as an open problem~\cite{codeharness}.
Using LLM reasoning as evidence for fault localization raises two problems.

\input{code/motivating-example}

\MyPara{Challenges.}
First, free-form reasoning cannot be executed.
A sentence such as ``the shifted values should stay non-positive'' does not say at which program point it applies, which version of a reassigned variable it refers to, or how to evaluate it.
It therefore cannot become a runtime check, and it cannot be linked to a statement.
Second, LLM reasoning is not reliable.
An LLM may state properties that are wrong or unrelated to the fault.
For the softmax function, ``the input should not be empty'' is true but unrelated to the fault, ``every returned value lies in $[0, 1]$'' describes a downstream symptom, and only the property of the shifted values describes the cause.
The LLM's confidence does not tell these cases apart: when our LLMs are asked directly for the faulty lines, they report a confidence of at least 0.9 for 76.2\% of their answers, but only 84.1\% of these answers are correct.
Running a property does not settle the question either, because passing tests can also violate a correct property when the fault executes without changing the output.

\MyPara{Our approach.}
We present \tool, which turns LLM reasoning into semantic constraints that can be executed and verified on the code graph (Figure~\ref{fig:design}).
To make the reasoning executable, \tool uses \emph{structured semantic grounding}.
It first builds a code graph of the target function from a lightweight static single assignment (SSA) form, whose anchors are function entries, returns, variable definitions and uses, branches, and loops, each with an unambiguous name.
The LLM then writes each semantic property as a constraint with a fixed structure: a category, an anchor on the code graph, and a Boolean Python expression over SSA-versioned variables.
\tool inserts each constraint as a runtime check at its anchor, runs the tests, and records which constraints each test violates.
The result is a \emph{semantic violation spectrum}, which takes the place of the coverage spectrum used by SBFL.
To make the reasoning verifiable, \tool uses \emph{counterfactual verification} (CF).
For a violated constraint, the LLM rewrites the anchored line so that the property holds, and \tool reruns the tests.
If the patch makes every failing test pass without breaking a passing test, the constraint and its line are marked Primary.
Verification thus confirms that a patch at the anchored line passes the tests, not that the constraint is true.
The LLM also lists candidate lines for the fault, and \tool applies the same check to the top one, which we call the \emph{hypothesis check}.
For fault localization, the final ranking lists verified lines first, then the other candidate lines, and then other lines where constraints are violated, so a developer can see which top-ranked lines the tests confirm.
\tool works at the function level: it takes a suspicious function and its tests, as given by a developer or by a function-level localizer~\cite{autofl,sievefl,farloc}.

In summary, we make the following contributions:
\begin{itemize}
  \item Structured semantic grounding, which turns LLM reasoning into constraints with fixed categories that are attached to anchors on a code graph and checked at runtime.
  \item \tool, which combines a semantic violation spectrum with counterfactual verification of constraint-guided patches and of the LLM's top answer, applied to fault localization.
  \item \sembugs, a dataset of 809 real bugs from 750 GitHub projects with executable tests, built from the cleaned release of TSSB-3M~\cite{tssb,ctssb}.
  \item An evaluation with four LLMs against LLM-based, coverage-based, reduction-based, and dynamic-invariant baselines, with a replication package.
\end{itemize}

\MyPara{Results.}
We evaluate \tool on \sembugs and on 33 QuixBugs programs~\cite{quixbugs}, and compare it with an LLM that sees only the function (\directllm), six SBFL formulas, delta debugging~\cite{zellerDD}, Daikon invariants~\cite{daikon}, and three ablations.
The constraints are executable: 99.4\% become runtime checks, and on the 185 programs in which SBFL ties the faulty line with at least half of the covered lines, their violations alone rank the faulty line first in 40.9\% of the cases, compared with 2.2\% for Ochiai.
To test whether the properties need an LLM, we replace the LLM constraints with the invariants that Daikon~\cite{daikon} infers from the passing tests; like a constraint, a Daikon invariant is a Boolean property of variables at a program point, so the same pipeline can ground, check, and score it.
The invariants rank the faulty line first in 3.8\% of the programs, and the LLM constraints in 32.2\%.
The constraints also make the LLM's answers verifiable: 68.4\% of \tool's top-1 answers carry a patch that the tests confirm, and 93.6\% of these are correct, whereas no threshold on the LLM's own confidence separates the answers as well.
\tool ranks the faulty line first in 81.4\% of the programs, compared with 82.0\% for an LLM given all of its inputs in one question, a difference that is not significant for any model, and 50.0\% for \directllm.

%% file: code/motivating-example.tex
\setminted{
  fontsize=\footnotesize,
  linenos,
  numbersep=6pt,
  xleftmargin=1.5em,
  frame=lines,
  framesep=2mm,
  % breaklines
}

\begin{listing}[tbp]
\caption{A buggy softmax function and its tests.}
\label{lst:softmax}

\begin{minted}[fontsize=\footnotesize, linenos]{python}
def softmax(xs):
    if not xs:
        return []
    m = max(xs)
    shifted = [x + m for x in xs]       # BUG: should be x - m
    exps = [exp(y) for y in shifted]    
    s = sum(exps)
    return [e / s for e in exps]        # inf/inf -> nan

# Test input 1: [0.0, 1.0]              # Passing
# Test input 2: [-1.0, 0.0, 1.0]        # Passing
# Test input 3: [1000.0, 1001.0]        # Failing
# Test input 4: [10000.0, 10001.0]      # Failing
\end{minted}

\end{listing}

%% file: sections/motivating.tex
\begin{figure}[t]
\centering
\includegraphics[width=\textwidth]{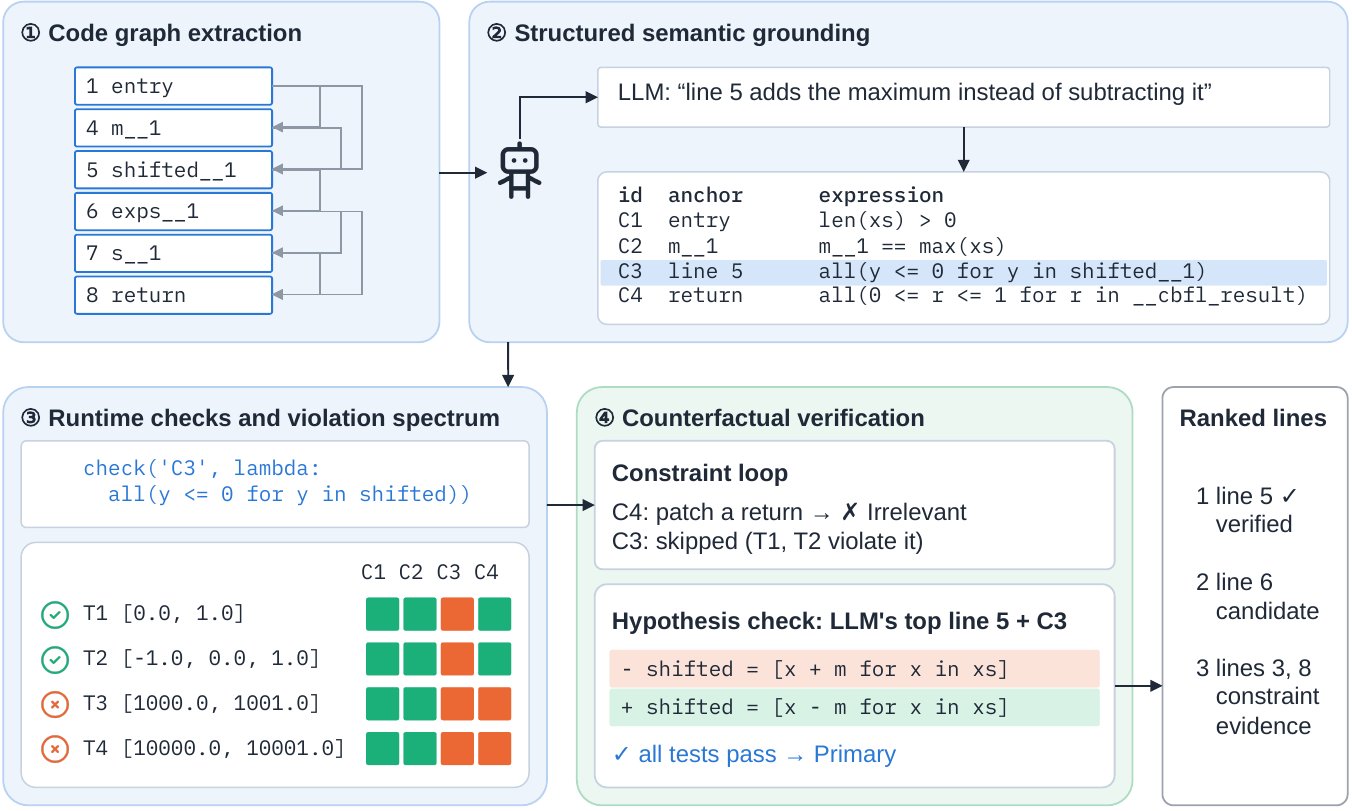}
\caption{\tool on the softmax function of Listing~\ref{lst:softmax}. Panels 1 to 4 correspond to Sections~\ref{sec:anchor} to~\ref{sec:counterfactual}. In panel 3, rows are tests, columns are constraints, and orange cells mark violations.}
\label{fig:approach}
\end{figure}

Listing~\ref{lst:softmax} shows a function that is meant to compute a numerically stable softmax.
Line~5 adds the maximum to each input instead of subtracting it.
Softmax does not change when all inputs are shifted by the same amount, so for small inputs such as \codefontsmall{[0.0, 1.0]} and \codefontsmall{[-1.0, 0.0, 1.0]} the function still returns the correct probabilities, and the first two tests pass.
For large inputs such as \codefontsmall{[1000.0, 1001.0]}, \codefontsmall{exp} overflows, the division yields \codefontsmall{nan}, and the last two tests fail.

\MyPara{Coverage and slicing.}
All four tests execute Lines~2 and~4 to~8 in the same order and with the same data dependences.
SBFL therefore assigns every executed line the same score, and a backward slice from the return value contains the whole computation.
What separates the passing from the failing tests is a semantic property of values: the shifted values should not be positive.
An LLM can state this property, but only as text, next to properties that are irrelevant or describe a symptom.

\MyPara{Constraints and verification.}
\tool asks the LLM for these properties in a fixed structure and attaches each one to the code graph.
The LLM returns four constraints:
$C_1$ at the function entry, \codefontsmall{len(xs) > 0};
$C_2$ after the definition of \codefontsmall{m\_\_1}, \codefontsmall{m\_\_1 == max(xs)};
$C_3$ after Line~5, \codefontsmall{all(y <= 0 for y in shifted\_\_1)};
and $C_4$ at the return, \codefontsmall{all(0 <= r <= 1 for r in \_\_cbfl\_result)}.
It also names Line~5 as its first candidate line.
Running the instrumented tests gives the semantic violation spectrum.
$C_1$ and $C_2$ hold in every test, and $C_4$ is violated by exactly the two failing tests.
$C_3$ is violated by all four tests, because the fault also executes in the passing tests.

The spectrum alone does not single out the cause.
$C_4$ separates failing from passing tests exactly, but it is checked at the return, after the fault; $C_3$ describes the fault, but because passing tests also violate it, it looks like an overly general constraint.
CF decides the case by patching and rerunning the tests.
The patch at a return for $C_4$ leaves the two failing tests failing, whereas the patch of Line~5 that $C_3$ suggests, to \codefontsmall{shifted = [x - m for x in xs]}, makes all four tests pass.
\tool therefore ranks Line~5 first and marks it as verified.
Section~\ref{sec:approach} follows this example through each stage.

%% file: sections/approach.tex
\tool takes a target function $P$ and a test suite $\mathcal{T} = \mathcal{T}^+ \cup \mathcal{T}^-$ of passing and failing tests.
It returns semantic constraints on the code graph of $P$, the outcome of every runtime check on every test, the lines and patches confirmed by CF, and, for fault localization, a ranked list of lines.
Figure~\ref{fig:design} shows the four stages.
We call the middle two stages, structured semantic grounding and the execution of runtime checks, \emph{semantic indexing}.

\subsection{Code Graph Extraction}\label{sec:anchor}

\tool parses $P$ with tree-sitter\footnote{\url{https://tree-sitter.github.io/tree-sitter/}} and derives a lightweight SSA form~\cite{cytronSSA} of it.
Each assignment defines a new version of its variable, written \codefontsmall{x\_\_2} for the second version of \codefontsmall{x}.
Augmented assignments such as \codefontsmall{b += 2} are lowered to plain assignments, so that the old and the new value have different names.
After an \codefontsmall{if} statement, \tool inserts a join assignment for every variable that is defined in a branch and read later, so each later read refers to exactly one version.
Every loop receives an integer identifier as a comment (\codefontsmall{\# loop\_\_id: 1}), which keeps nested loops apart.
The rewriting records a definition map from each version to its source line and a use map from each version to the statements that read it.

The code graph consists of the anchors that this form makes explicit: the function entry, every return, every definition and read of an SSA version, the end of every branch of an \codefontsmall{if}, the head and tail of every loop, every call site, and every source line.
Its edges link each SSA definition to its reads.
The SSA form is used in two ways: it gives the LLM unambiguous variable names, and it gives the instrumenter the exact position of each anchor.
The program that runs is the original one, instrumented at the anchors (Section~\ref{sec:spectrum}); \tool does not execute the SSA form.

\MyPara{Example.}
Panel~1 of Figure~\ref{fig:approach} shows the code graph of the softmax function, with the SSA versions defined on Lines~4 to~7 and the links from each definition to its reads.

\subsection{Structured Semantic Grounding}\label{sec:inference}

\input{tables/fse_schema}
\input{code/prompt-template}

\tool asks the LLM for semantic properties in a closed schema (Table~\ref{tab:schema}).
A constraint is a tuple $C = \langle \mathit{id}, \kappa, \rho, \alpha, \varphi, \iota \rangle$.
The category $\kappa$ is one of nine labels.
Six of them are checked at one program point, and three temporal ones compare states across points, such as the state at entry and at exit of a call.
The region $\rho$ says where the check runs, and the anchor $\alpha$ selects the code-graph element, for example the SSA version \codefontsmall{shifted\_\_1} for an \texttt{AFTER\_DEF} constraint.
The specification $\varphi$ is a Boolean Python expression over the variables visible at the anchor; the prompt suggests comparisons, arithmetic, and the built-ins \codefontsmall{len}, \codefontsmall{abs}, \codefontsmall{all}, \codefontsmall{any}, and \codefontsmall{sum}, but any expression that parses is accepted.
The intent $\iota$ is one sentence of explanation.
With a fixed set of labels, a fixed set of handlers can place and run every constraint.

Listing~\ref{lst:prompt} shows the prompt in condensed form.
It contains the target function with line numbers, its SSA form with the definition map, the passing tests, and the failing tests with their error messages.
In this \emph{inference call}, the LLM returns, in one JSON object, a fault hypothesis, a fault line, a list of up to five candidate lines in order of suspicion that starts with the fault line, and the constraints.
The prompt asks for one \texttt{LINE} constraint at the fault line and for at least two constraints near it at fine-grained anchors such as definitions, uses, and branches, and it limits constraints at the entry and the return to two, because those regions cannot point at a statement.
Constraints must use SSA names for variable anchors, so a constraint on \codefontsmall{x\_\_2} cannot be confused with one on \codefontsmall{x\_\_1}.

Grounding binds each constraint to its insertion sites.
\tool resolves an SSA version through the definition map or the use map, a loop identifier through the loop comments, and a branch through its \codefontsmall{if} line.
An anchor with several sites, such as a variable read in both branches of an \codefontsmall{if}, receives one check per site.
The grounding rules also handle common ways in which LLM output deviates from the schema.
For example, a dotted name such as \codefontsmall{self.count} is grounded after the assignments to that expression, a version number that does not exist is grounded at every definition of the variable, and a \texttt{LINE} anchor on a continuation line is moved to the enclosing statement.
Constraints that cannot be grounded, such as those that refer to a variable absent from the function or whose expression is not valid Python, are dropped.

\MyPara{Example.}
For Listing~\ref{lst:softmax}, the LLM states the fault hypothesis that Line~5 adds the maximum instead of subtracting it, names Lines~5 and~6 as candidate lines, and returns the four constraints of Section~\ref{sec:motivation}.
The constraint on the shifted values, for instance, is the \texttt{LINE} constraint at the fault line, $C_3 = \langle$C3, \texttt{VALUE\_RANGE}, \texttt{LINE}, 5, \codefontsmall{all(y <= 0 for y in shifted\_\_1)}, $\iota_3\rangle$, where $\iota_3$ reads ``after subtracting the maximum, no value is positive, so \codefontsmall{exp} cannot overflow''.
Panel~2 lists the grounded constraints: the definition map resolves $C_2$ to Line~4, $C_3$ is checked after Line~5, and $C_4$ at both returns.

\subsection{Runtime Checks and Semantic Violation Spectrum}\label{sec:spectrum}

\tool inserts each grounded constraint as a runtime check \codefontsmall{check(cid, lambda: $\varphi$)} at its sites in the original program (Figure~\ref{fig:approach}, panel~3).
Inside the check, each SSA name is replaced by its source variable, which holds the value of that version at the anchor; the check for $C_3$, for example, reads \codefontsmall{shifted}.
If a name is not bound yet, the check raises an exception and the program itself continues unchanged.
Each \codefontsmall{return X} is rewritten to bind its value to \codefontsmall{\_\_cbfl\_result} before the checks of the return run, so output properties can refer to it.
Checks that would follow a \codefontsmall{raise}, \codefontsmall{break}, or \codefontsmall{continue} are placed before it, and a check on a line that holds other code is appended to that line.
A snapshot check (\texttt{ON\_EXIT}) compares values stored at entry with the state before each return, and temporal constraints keep per-test state in context variables.
A pytest plugin resets this state before each test and writes the violations after it.
A check is violated when $\varphi$ evaluates to false or raises an exception; \tool records which of the two happened.

The violations form the semantic violation spectrum $V \in \{0,1\}^{|\mathcal{T}| \times n}$, where $V_{ij} = 1$ if test $t_i$ violates constraint $C_j$.
As in SBFL~\cite{tarantula,OchiaiSbflCov}, \tool counts for each constraint the failing tests that violate it ($e_f$) and the passing tests that violate it ($e_p$), and scores it with the Ochiai coefficient
\[
  \sigma(C_j) = \frac{e_f(C_j)}{\sqrt{|\mathcal{T}^-| \cdot \big(e_f(C_j) + e_p(C_j)\big)}},
\]
with $\sigma(C_j) = 0$ if no test violates $C_j$.
A constraint passes its score to the lines of its anchor.
Coarse regions receive lower weights, because they confirm that the function is wrong without pointing at a statement: \texttt{ENTRY} 0.2, \texttt{ANY\_RETURN} and \texttt{EXIT} 0.3, \texttt{ON\_EXIT} 0.5, \texttt{LINE} 0.6, loop and call anchors 0.9, and the remaining fine-grained anchors 1.0.
A \texttt{LINE} constraint also passes 55\% of its weighted score to the adjacent lines and 25\% to the lines two away.
The score of a line is the maximum weighted score of the constraints anchored there; we call these scores the constraint evidence of the line.

\MyPara{Example.}
Panel~3 shows the spectrum: $C_3$ is violated by all four tests, and $C_4$ by the two failing tests, whose outputs are \codefontsmall{nan}.
With $|\mathcal{T}^-| = 2$, this gives $\sigma(C_1) = \sigma(C_2) = 0$, $\sigma(C_3) = 2/\sqrt{2 \cdot 4} \approx 0.71$, and $\sigma(C_4) = 2/\sqrt{2 \cdot 2} = 1$.
$C_4$ has the highest score but only confirms that the output is wrong; with weight 0.3, it gives Lines~3 and~8 a score of 0.3, while $C_3$, with weight 0.6, gives Line~5 a score of 0.42 and the lines around it less.
Line~5 thus gets the most constraint evidence, from a constraint that passing tests also violate.

\subsection{Counterfactual Verification}\label{sec:counterfactual}

\input{algos/fse_cf_ranking}

A violated constraint is only a hypothesis about the fault, and the spectrum cannot always distinguish a cause from a symptom or from an overly general constraint (Section~\ref{sec:motivation}).
CF tests each hypothesis by patching the code (Algorithm~\ref{algo:cf}).
For a constraint $C_j$ anchored at line $\ell_j$, the LLM receives the constraint expression, its intent, the statement at $\ell_j$, and the function, and returns one replacement for that statement.
\tool applies the replacement, reruns all tests, and classifies the outcome.
The patch is \emph{Primary} if every failing test passes and no passing test fails, \emph{Secondary} if some but not all failing tests pass and no passing test fails, \emph{Regressing} if a failing test passes but a passing test fails, \emph{Irrelevant} if no failing test passes, and \emph{PatchInvalid} if the patched program does not parse, import, or run.

CF runs in two parts.
The constraint loop visits up to five constraints with nonzero scores, in the order of their scores.
To save patch requests, it skips constraints that a passing test violates and constraints without an anchor line.
It stops at the first Primary patch, and \tool attributes that patch to the lines whose tokens it changes.
A patch replaces the whole statement that contains the anchor line, or only the header of a compound statement such as an \codefontsmall{if}, so that it never removes a block body, and a multi-line patch is re-indented to the statement's indentation.
The hypothesis check then patches the LLM's top candidate line, if there is one, guided by the constraint anchored at that line (a \texttt{LINE} constraint first) or, if there is none, by the fault hypothesis.
It runs regardless of the spectrum score, so a correct constraint that passing tests violate can still confirm the answer.

\MyPara{Example.}
The constraint loop visits $C_4$ and $C_3$, the two constraints with nonzero scores, and skips $C_3$ because the passing tests violate it.
For $C_4$, the patch at the return does not make a failing test pass, so it is Irrelevant.
The hypothesis check then patches the top candidate, Line~5, guided by $C_3$; the patch makes all four tests pass and is Primary (panel~4).

\subsection{Ranking for Fault Localization}\label{sec:ranking}

For fault localization, \tool orders lines in three tiers, and a later tier only adds lines that are not ranked yet.
The first tier holds verified lines.
If the hypothesis check is Primary, it holds the top candidate line; otherwise it holds the lines changed by a Primary constraint patch, in the order of the candidate list.
CF thus changes the LLM's ranking only when its first candidate fails verification.
The second tier holds the candidate lines in the LLM's order, and the third holds the remaining lines with constraint evidence, ordered by score.
Lines with equal scores are tied.
Removing CF leaves candidate lines followed by constraint evidence, which is the ablation \toolnocf in Section~\ref{sec:evaluation}.

\MyPara{Example.}
For Listing~\ref{lst:softmax}, the first tier holds Line~5, verified by the hypothesis check.
The second tier adds the other candidate, Line~6, and the third adds Lines~3 and~8, tied at 0.3, then Lines~4 and~7.
The developer sees Line~5 first, together with the patch that makes all tests pass and the constraint $C_3$ that explains the fault.

\subsection{Implementation}

\tool is implemented in Python on tree-sitter and pytest.
It calls LLMs through an OpenAI-compatible API or through a command-line coding agent run as a single-turn text model in an empty, read-only directory.
Tests run in separate processes with a limit of 10 seconds per test and a memory limit.
CF uses the same LLM, at temperature 0 where it can be set, and runs each patched program in a temporary copy without runtime checks, so that no check can affect CF.

%% file: tables/fse_schema.tex
\begin{table}[t]
\centering
\caption{Constraint schema: categories (left) and anchor regions (right).}
\label{tab:schema}
\footnotesize
\setlength{\tabcolsep}{3.5pt}
\resizebox{\linewidth}{!}{%
\begin{tabular}[t]{@{}ll@{}}
\toprule
Category & Property \\
\midrule
\texttt{PRECONDITION} & valid inputs \\
\texttt{POSTCONDITION} & valid outputs \\
\texttt{VALUE\_RANGE} & bounds of a value \\
\texttt{RELATION} & relation between values \\
\texttt{DERIVED\_CONSISTENCY} & derived values agree \\
\texttt{INVARIANT\_LOOP} & holds in every iteration \\
\texttt{TEMPORAL\_CALL\_SNAPSHOT} & entry vs.\ exit state \\
\texttt{TEMPORAL\_UNTIL\_OVERWRITTEN} & last write wins \\
\texttt{TEMPORAL\_RESOURCE\_LIFETIME} & acquire--release pairs \\
\bottomrule
\end{tabular}\hspace{0.8em}%
\begin{tabular}[t]{@{}lll@{}}
\toprule
Region & Anchor & Check position \\
\midrule
\texttt{ENTRY} & -- & function entry \\
\texttt{ANY\_RETURN}, \texttt{EXIT} & -- & before each return \\
\texttt{AFTER\_DEF} & SSA variable & after its definition \\
\texttt{BEFORE\_USE} & SSA variable & before each read \\
\texttt{AFTER\_BRANCH} & \texttt{if} line, branch & end of the branch \\
\texttt{LOOP\_HEAD}, \texttt{LOOP\_TAIL} & loop id & start, end of an iteration \\
\texttt{AFTER\_CALL} & callee & after the call \\
\texttt{LINE} & line number & after the statement \\
\texttt{ON\_EXIT} & -- & entry and before each return \\
\bottomrule
\end{tabular}%
}
\end{table}

%% file: code/prompt-template.tex
\begin{listing}[t]
\caption{Condensed constraint inference prompt.}
\label{lst:prompt}
\begin{minted}[fontsize=\scriptsize, linenos, breaklines, breaksymbolleft={}, frame=lines]{text}
Given a buggy Python function and passing/failing tests: identify the line most
likely to contain the bug, and place semantic constraints that are true on
passing tests and false on failing tests. Output only JSON:
{ "fault_hypothesis": "<what is wrong, on which line>",
  "fault_line": <int>, "fault_lines": [<up to 5 lines, most suspicious first>],
  "constraints": [ { "id": "C1", "category": "<one of 9 categories>",
      "instrument": { "region": "<one of 11 regions>", "anchor": {...} },
      "spec": { "expr": "<Boolean expression over SSA names>" },
      "intent": "<one sentence>", "confidence": <0..1> } ] }
Rules: one LINE constraint at fault_line; at least two AFTER_DEF, BEFORE_USE,
AFTER_BRANCH or LOOP_TAIL constraints; at most two ENTRY/ANY_RETURN/EXIT ones.
Anchors: AFTER_DEF/BEFORE_USE {"var": "x__2"}; LOOP_* {"loop_id": N}; LINE {"line": N}.
Expressions: comparisons, arithmetic, len/abs/all/any/sum; __cbfl_result = return value.
### Program (numbered)      ### SSA form with def map (m__1 -> 'm', line 4)
### Passing tests           ### Failing tests with errors
\end{minted}
\end{listing}

%% file: algos/fse_cf_ranking.tex
\begin{algorithm}[t]
\small
\DontPrintSemicolon\SetAlgoNoEnd
\SetKwInOut{Input}{Input}\SetKwInOut{Output}{Output}
\SetKwFunction{Run}{RunTests}\SetKwFunction{Patch}{LLMPatch}\SetKwFunction{Classify}{Classify}\SetKwFunction{Changed}{ChangedLines}
\Input{function $P$; tests $\mathcal{T}$ with failing tests $\mathcal{T}^-$; constraints $C_1..C_n$ with spectrum scores $\sigma$ and anchor lines $\ell$; candidate lines $L$; constraint line scores $S$}
\Output{ranked lines}
\BlankLine
$\mathit{verified} \leftarrow \emptyset$\;
\ForEach(\tcp*[f]{constraint loop}){$C_j$ in the top five constraints by $\sigma_j > 0$}{
  \lIf{$C_j$ is violated by a passing test \textbf{or} has no anchor line}{\textbf{continue}}
  $P' \leftarrow P$ with the statement at $\ell_j$ replaced by \Patch{$C_j$, $P$}\;
  $s_j \leftarrow$ \Classify{\Run{$P'$, $\mathcal{T}$}}\tcp*[r]{Primary, Secondary, Regressing, Irrelevant, PatchInvalid}
  \lIf{$s_j = \textsc{Primary}$}{$\mathit{verified} \leftarrow$ \Changed{$P$, $P'$}; \textbf{break}}
}
\If(\tcp*[f]{hypothesis check}){$L$ is not empty}{
  $C_h \leftarrow$ the constraint anchored at $L[0]$, or the fault hypothesis if there is none\;
  $P'' \leftarrow P$ with the statement at $L[0]$ replaced by \Patch{$C_h$, $P$}\;
  \lIf{\Classify{\Run{$P''$, $\mathcal{T}$}} $= \textsc{Primary}$}{$\mathit{verified} \leftarrow \{L[0]\}$}
}
\Return $\mathit{verified}$ (in the order of $L$), then $L$, then $S$ by score, each line listed once\;
\caption{Counterfactual verification and final ranking.}
\label{algo:cf}
\end{algorithm}

%% file: sections/evaluation.tex
We evaluate \tool on fault localization with four research questions.

\begin{itemize}
  \item[\textbf{RQ1}] \emph{Fault localization.} What do executable and verifiable constraints add to fault localization, compared with LLM-based and coverage-based baselines?
  \item[\textbf{RQ2}] \emph{Constraints and verification.} Are the constraints executable, and what does CF confirm?
  \item[\textbf{RQ3}] \emph{Source of constraints.} How do LLM constraints compare with dynamic invariants as a source of semantic properties?
  \item[\textbf{RQ4}] \emph{Cost.} What does \tool cost compared with a single LLM question?
\end{itemize}

\subsection{Datasets}\label{sec:datasets}

\input{tables/fse_funnel}

\sembugs contains real bugs from open-source Python projects, each with its fault in an expression.
The buggy and the fixed function are taken verbatim from a fix commit, the ground truth is computed from the diff, and only the harness that makes a function run on its own is generated.
Table~\ref{tab:funnel} shows how many fixes remain after each construction step.

\input{tables/fse_rq1}

\MyPara{Mining and filtering.}
We start from CTSSB-1M, the cleaned set of 966,985 bug fixes released with TSSB-3M~\cite{tssb,ctssb}.
We keep the 13 patterns that change an expression: a binary or unary operator, a binary operand, a numeric or Boolean literal, a constant type, an identifier, an attribute, a keyword argument, a called function, swapped arguments, and a condition that is more or less specific than needed.
We further require that the fix lies inside a function of a non-test source file, that its faulty location is unambiguous, and that it comes from a non-fork repository with at least 100 stars, which we check through the GitHub API.
From the 18,253 fixes that remain, we sample 3,322, at most 300 per pattern and at most two per project, so that no pattern or project dominates, and download both versions of the file.
A structural filter then keeps the fixes whose buggy and fixed functions contain the same multiset of abstract syntax tree (AST) statement types.
The fault therefore lies in an expression and does not change the statement structure or the control-flow graph of the function, which is the setting in which syntactic spectra give little signal.

\MyPara{Semantic filter.}
CTSSB-1M labels a commit as a bug fix with a heuristic on its message, so many fixes change a requirement instead of repairing a fault.
For such fixes the correct value cannot be inferred from the program, and no fault localization technique can be expected to find them.
An LLM judge (Claude Opus 5.5, one turn, no tools) sees both functions, the diff, the docstring, and the commit message, and keeps a fix only if it repairs a defect whose correct behavior follows from the code; two mechanical checks also drop fixes that change only whitespace or string literals.
This removes 982 fixes: 324 changed requirements, 191 adaptations to external changes, 157 defects whose fix cannot be inferred, and 310 text, formatting, or refactoring changes.

\MyPara{Harness generation.}
For each of the 1,373 remaining fixes, an LLM coding agent (Claude Code, mostly with Claude Opus 5.5) writes a self-contained module with minimal stubs for the external names the function uses, and a pytest file with 15 to 20 tests.
The agent works in an isolated temporary directory that contains only a task description and the fixed file.
It never sees the buggy version, the diff, the bug pattern, or the commit message, so the tests describe the intended behavior rather than the bug.
A hook in the agent runner denies file access outside the directory and all network and package-installation commands.
The buggy module is then derived mechanically by putting the buggy line back into the self-contained fixed module, so both versions share the same stubs and tests.
The agent produces a harness for 1,344 of them (97.9\%).

\begin{figure}[t]
\centering
\begin{minipage}[t]{0.49\textwidth}
  \centering
  \includegraphics[width=\linewidth]{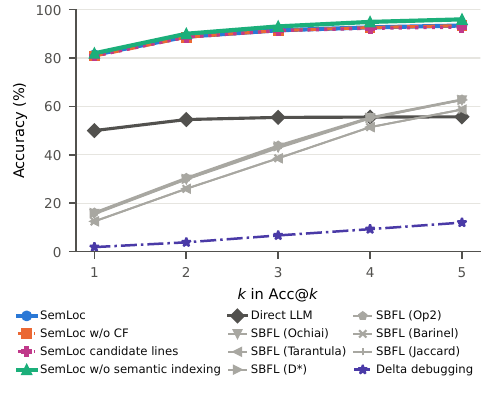}
  \caption{Acc@$k$ on \sembugs, all models. SBFL formulas in gray; Tarantula and Barinel coincide.}
  \label{fig:acck}
\end{minipage}\hfill
\begin{minipage}[t]{0.49\textwidth}
  \centering
  \includegraphics[width=\linewidth]{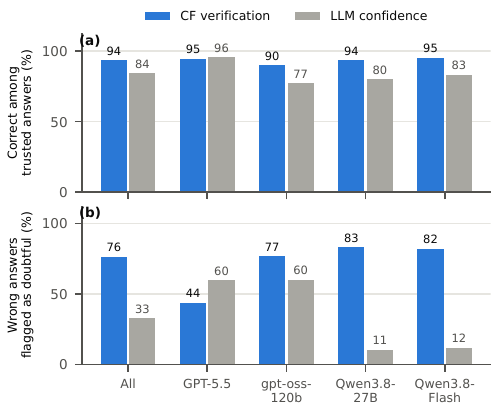}
  \caption{Trust signals for the top-1 answer on \sembugs: (a) share of trusted answers that are correct; (b) share of wrong answers flagged as doubtful. CF verification applies to \tool, LLM confidence (trusted if at least 0.9) to \toolnoidx.}
  \label{fig:verified}
\end{minipage}
\end{figure}

\MyPara{Verification.}
We run every instance three times and keep it only if the fixed version passes every test, the buggy version fails at least one test and passes at least three, and the outcomes are identical across the three runs.
This step removes 293 instances whose tests do not expose the bug reliably.
Removing near-duplicates, with a token similarity above 0.9 or the same repository and edit, removes 7 more.
We also record the per-test line coverage of the buggy version for SBFL and the statistics below.

\MyPara{Failure mode.}
Finally, we exclude the 235 instances in which a failing test crashes on the faulty line itself, for example with an \codefontsmall{AttributeError} raised by a misspelled attribute, because the stack trace already names the line.
In the remaining instances, a wrong value flows on and is detected later by an assertion, an exception elsewhere, or a timeout.
This leaves the 809 programs of \sembugs.

\MyPara{Statistics.}
Table~\ref{tab:datasets} summarizes \sembugs.
Its programs come from 750 repositories and commits from 2005 to 2021, and a function has 35.5 lines, 19.5 tests, and 5.1 failing tests on average.
In 185 programs (22.9\%), at least half of the function's lines that some test covers, the faulty line included, are covered by as many failing and as many passing tests as the faulty line, so every SBFL formula ties them; we call this \emph{coverage collapse}.
We also use QuixBugs~\cite{quixbugs}, a benchmark of classic algorithm implementations with known bugs and its own test cases.
We keep the 33 Python programs that pass the same verification step, with at least one passing test instead of three.

\subsection{Models, Baselines, and Metrics}

\MyPara{LLMs.}We run every LLM-based method with four LLMs.
GPT-5.5 runs three times through the Codex command-line agent\footnote{\url{https://github.com/openai/codex}} with its default temperature.
gpt-oss-120b, Qwen3.8-27B, and Qwen3.8-Flash run through an OpenAI-compatible API at temperatures 0.0, 0.3, and 0.8, with thinking disabled for the Qwen models.
This gives 12 runs per dataset, each with 24 parallel workers.

\MyPara{Baselines.}We compare \tool with three ablations and four baselines.
\toolnocf ranks the candidate lines and then the constraint evidence, without CF, and \toolcand ranks only the candidate lines.
\toolnoidx removes semantic indexing: it is a separate LLM call that receives all of \tool's inputs, namely the target function with line numbers, its SSA form, the passing tests, and the failing tests with their errors, together with the plain instruction to return the five most suspicious lines.
It writes no constraints and executes nothing.
\directllm receives only the target function with line numbers.
SBFL uses line coverage of the same tests with six formulas: Ochiai~\cite{OchiaiSbflCov}, Tarantula~\cite{tarantula}, D*~\cite{dstar}, Op2~\cite{op2}, Barinel~\cite{barinel}, and Jaccard~\cite{jaccardAbreu2007}.
Delta debugging (DD)~\cite{zellerDD} removes statements of the target function and keeps the smallest set for which every failing test still fails and every passing test still passes; we use the ddmin algorithm~\cite{ddmin} with at most 80 test runs, 120 seconds per test run, and 15 minutes per program, and rank the remaining statements as ties.
DD removes simple statements such as assignments, calls, and returns, but not the conditions of \codefontsmall{if} or \codefontsmall{while} statements.
For RQ3, we use Daikon~\cite{ernstDaikon2001,daikon} and the invariant annotator of The Debugging Book~\cite{debuggingbook}, described in Section~\ref{sec:rq3}.

\MyPara{Metrics.}We report Acc@$k$ for $k \in \{1, 3, 5\}$, the share of programs whose faulty line is among the first $k$ ranked lines.
Tied lines take the worst rank of their group, so a method cannot gain from ties.
We also report \pctflag, the share of the target function's executable lines that a method ranks at all; lower values mean fewer lines to inspect.
For LLM-based methods we average over the runs of each model.
To compare two methods, we use the exact McNemar test~\cite{mcnemar} over \emph{program runs}, pairs of a program and a run, with the runs of a model pooled.
Because the runs of one program are not independent, we also bootstrap the per-program differences over programs (2,000 samples), and call a difference significant only if $p < 0.05$ and the 95\% interval excludes zero.

\input{tables/fse_patterns}

\subsection{RQ1: Fault Localization}\label{sec:rq1}

\MyPara{Setup.}
We run \tool, the three ablations, and \directllm with every model on both datasets, and SBFL once per dataset.
Besides the full \sembugs, we report the programs with coverage collapse, where syntactic spectra give little signal.
We compare \tool with every other method by the McNemar test.
Finally, we ask whether a developer can tell if the first answer is right.
We call a top-1 answer \emph{verified} if CF confirms it, that is, if a patch at that line makes every failing test pass without breaking a passing test, and \emph{unverified} otherwise; as for Acc@1, an answer is correct if it is the faulty line.
As a comparison, we use the confidence between 0 and 1 that the LLM reports for its answer in \toolnoidx, and trust an answer if this confidence is at least 0.9; \tool's inference call reports a confidence only per constraint, which \tool does not use.

\MyPara{Comparison with baselines.}
Table~\ref{tab:rq1} reports the accuracy of all methods, and Figure~\ref{fig:acck} plots it for $k$ from 1 to 5.
On \sembugs, \tool reaches 81.4\% Acc@1, 91.5\% Acc@3, and 93.4\% Acc@5.
\directllm reaches 50.0\%, 55.5\%, and 55.7\%, and \tool is significantly more accurate for every model and every $k$.
SBFL is much less accurate: with each of the six formulas, it fails to rank the faulty line first in at least 83.8\% of the programs and misses it in the top five in at least 37.1\%.
DD ranks the faulty line first in 1.9\% of the programs and within five in 12.0\%: it keeps 12.9 statements on average, and because it cannot remove conditions, its set contains the faulty line in only 2 of the 170 condition bugs.
On the 185 programs with coverage collapse, Ochiai misses the faulty line in the top five in 55.1\% of the cases, while \tool places it first in 86.3\% and in the top five in 96.5\%.
\tool also ranks fewer lines: 35.5\% of the executable lines, compared with 63.1\% for SBFL and 49.0\% for DD.
\directllm ranks fewer lines (9.8\%) but misses the fault far more often.

\input{tables/fse_rq2}

\MyPara{Comparison with \toolnoidx.}
\toolnoidx is a strong ablation: it gives the LLM every input that \tool uses and asks a single question for the five most suspicious lines.
At the first rank, the accuracy difference is not significant for any model (81.4\% vs.\ 82.0\%).
Within five lines, the single question is more accurate (96.0\% vs.\ 93.4\% Acc@5), significantly for three models; for GPT-5.5 the difference is not significant at any $k$, and on QuixBugs \tool is ahead at Acc@1 and Acc@3.
The single question, however, returns only line numbers and a confidence value, whereas \tool's ranking rests on runtime checks and, for two thirds of its first answers, on a patch that the tests confirm.

\MyPara{Verified answers.}
Figure~\ref{fig:verified} compares the two signals on two measures: the share of trusted answers that are correct (Figure~\ref{fig:verified}a) and the share of wrong answers that are flagged as doubtful (Figure~\ref{fig:verified}b).
CF verifies 68.4\% of \tool's top-1 answers, and 93.6\% of the verified answers are correct, between 89.9\% and 95.4\% per model; for 79.4\% of the correct ones, the verifying patch is the developer's fix (Section~\ref{sec:rq2}).
In addition, 76.5\% of \tool's wrong top-1 answers are unverified: a developer who inspects only the unverified answers, about a third of all, sees three quarters of the wrong ones.
LLM confidence separates the answers much less well: the LLM reports a confidence of at least 0.9 for 76.2\% of its answers, 84.1\% of these trusted answers are correct, and only 33.1\% of its wrong answers fall below the threshold.
No threshold matches CF: a threshold of 0.96 trusts 14.5\% of the answers at 93.1\% precision, and higher ones reach 93.6\% only for fewer than 10\% of them.
Only for GPT-5.5 is confidence as informative as CF: 95.6\% of its confident answers are correct (CF: 94.7\%), and it flags 59.7\% of the wrong answers (CF: 43.8\%).

\MyPara{Contribution of constraint evidence and CF.}
\toolcand reaches 80.8\%, 91.3\%, and 92.7\%; the constraint evidence raises Acc@5 to 93.4\% (\toolnocf), and CF raises Acc@1 to 81.4\%.
CF changes the ranking only when the LLM's first answer fails the hypothesis check and a constraint patch succeeds elsewhere; at Acc@1 it gains 25 program runs and loses 5 for gpt-oss-120b, 19 vs.\ 3 for Qwen3.8-27B, 19 vs.\ 4 for Qwen3.8-Flash, and 21 vs.\ 20 for GPT-5.5.
Against \toolcand, \tool is significantly more accurate at every $k$ for gpt-oss-120b and Qwen3.8-27B and at Acc@1 for Qwen3.8-Flash, but not for GPT-5.5.

\MyPara{Bug patterns.}

By bug pattern (Table~\ref{tab:patterns}), every LLM-based method is least accurate at Acc@1 on the two condition patterns, where \directllm reaches only 21.9\% and 24.5\%, SBFL at most 5.7\% and 1.2\%, and DD 0.0\%.
A wrong condition sends execution into the wrong branch, and the symptom appears later, in that branch or at the return.
Checks at branch ends and patches of the condition let \tool trace the symptom back to the condition, as in \texttt{keepCallout} (Section~\ref{sec:rq2}); when the condition is less specific than needed, \tool has the highest Acc@1 (65.1\%).

\MyPara{QuixBugs.}
With the tests that come with QuixBugs, \tool reaches 74.2\% Acc@1 and 96.0\% Acc@5, compared with 72.2\% and 98.7\% for \toolnoidx, 71.0\% and 76.8\% for \directllm, and 24.2\% and 54.5\% for Ochiai.
CF verifies 64.9\% of \tool's top-1 answers, and 91.1\% of these are correct, compared with 74.4\% of the trusted answers of \toolnoidx.

\rqanswer{1}{\tool ranks the faulty line first in 81.4\% of the \sembugs programs, with no significant difference from an LLM given the same inputs, and far more often than \directllm and SBFL, also where coverage gives little signal.
Unlike the single question, it shows which first answers the tests confirm: two thirds carry a verified patch, 93.6\% of which are correct.}

\subsection{RQ2: Semantic Constraints and Counterfactual Verification}\label{sec:rq2}

\begin{figure}[t]
\centering
\begin{minipage}[c]{0.5\textwidth}
\includegraphics[width=\linewidth]{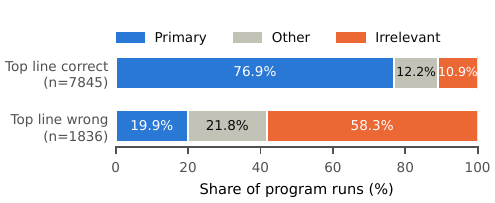}
\end{minipage}\hfill
\begin{minipage}[c]{0.46\textwidth}
\caption{Outcome of the hypothesis check on \sembugs, by whether the LLM's top candidate line is the faulty line. Other: Secondary, Regressing, or PatchInvalid; the 27 program runs without a candidate line are left out.}
\label{fig:hypothesis}
\end{minipage}
\end{figure}

\MyPara{Setup.}
We analyze all constraints of the RQ1 runs on \sembugs.
A constraint is instrumented if it is grounded and inserted without error, and executed if at least one test evaluates it.
We count the constraints that at least one failing test violates and those that at least one passing test violates.
To check whether a constraint describes intended behavior, we instrument the fixed program with the same constraints and run the same tests; a constraint holds on the fixed program if at least one test evaluates it and no test violates it.
For CF, we count the Primary patches of the constraint loop and the share changing the faulty line, and we split the outcomes of the hypothesis check by whether the top candidate line is the faulty line.
We leave out QuixBugs, whose fixed programs are not line-aligned with the buggy ones.
Table~\ref{tab:rq2} follows the constraints from generation to verification.

\MyPara{Executable constraints.}
The LLMs write between 3.01 and 4.13 constraints per program.
Of all constraints, 99.4\% are grounded and instrumented, and 94.0\% are executed by at least one test; the remaining ones sit on paths that no test reaches.
Also, 36.0\% of the constraints call functions beyond the suggested built-ins, most often \codefontsmall{isinstance} and \codefontsmall{hasattr}.
Compared with the program without checks, the checks change only 0.11\% of the test outcomes.

\MyPara{Checks written by the LLM.}
The LLM could also write the checks into the code itself.
The prompt would have to state the insertion rules, and the LLM would have to follow them for every constraint: record a violation instead of ending the test, bind the return value, place checks before a \codefontsmall{raise} or \codefontsmall{break}, and keep per-test state for temporal constraints.
\tool implements these rules once, so they always hold, and leaves the rest of the program unchanged.
Breaking the first rule changes the tests: in 61.2\% of the program runs a passing test violates a constraint, and in 11.8\% every passing test does, so with \codefontsmall{assert} statements the spectrum would lose these tests.
Rewriting code is also error-prone: even for one statement in CF, 5.3\% of the patches do not parse, import, or run.

\MyPara{Correct and wrong constraints.}
Between 43.8\% and 67.3\% of the constraints are violated by at least one failing test, so most constraints react to the fault.
A violation alone, however, does not show whether a constraint states the intended behavior.
We check this on the fixed program: a constraint that holds there is consistent with the correct program, and one that is violated there states a wrong property.
Crossing the two questions splits the executed constraints, not all as in Table~\ref{tab:rq2}, into four kinds, which the \texttt{heapify} function of \sembugs illustrates; its Line~2, \codefontsmall{p = (size/2)-1}, uses float division where the fix uses \codefontsmall{size//2}.
First, 27.4\% hold on the fixed program and are violated by failing tests, such as \codefontsmall{p\_\_1 == size//2 - 1}: they state the intent that the fault breaks.
Second, 32.0\% hold but are never violated by a failing test, such as the loop invariant \codefontsmall{p < size}: true, but unrelated to the fault.
Third, 29.5\% are violated by failing tests although they are wrong, such as \codefontsmall{p\_\_1 != size/2 - 1}, which restates the buggy expression instead of the intent and also fails on the fixed program for every even \codefontsmall{size}; the remaining 11.0\% are wrong and never violated.

\input{tables/fse_regions}
\input{code/case-keepcallout}

On the buggy program, the spectrum cannot tell the first kind from the third: both are violated by failing tests, and in \texttt{heapify} the wrong constraint is violated by 17 failing tests and the correct one by only 9.
Only the fixed program separates them, and it is not available when a fault is being localized.
Of the constraints that failing tests violate, 48.1\% state the intended behavior and 51.9\% do not, so about half of the spectrum's evidence comes from wrong properties.
CF does not depend on whether a constraint is right: it uses the constraint only to choose a line and to guide a patch, and the tests decide.
In \texttt{heapify}, the patch guided by the wrong constraint changes Line~2 to \codefontsmall{p = (size//2)-1}, and all tests pass.
Primary patches are anchored at the faulty line in 90.7\% of the cases when their constraint holds on the fixed program and in 89.5\% when it does not.

\MyPara{Verified patches.}
The constraint loop produced 2,831 Primary patches, and 90.0\% of them change the faulty line (between 86.6\% and 93.0\% per model).
Figure~\ref{fig:hypothesis} shows the hypothesis check.
When the LLM's top candidate line is correct, CF verifies it in 76.9\% of the program runs and finds it Irrelevant in 10.9\%.
When the top line is wrong, CF verifies it in only 19.9\% and finds it Irrelevant in 58.3\%.
The outcome of the check thus separates correct from wrong answers, which is the basis of the verified-answer precision in RQ1.
A wrong line can still pass the check, because the tests are the only oracle and patches at more than one place can often make all tests pass.
Of the 366 wrong top lines that pass the check, 55.5\% are within one line of the faulty line (median distance 1), compared with 26.5\% of the wrong lines that CF finds Irrelevant (median distance 3).
More than half of the verified wrong answers are thus test-passing patches next to the faulty line.
Most verified patches are the developer's fix, with the same AST as the fixed program: 79.8\% of the Primary patches of the hypothesis check when the top line is correct, and 69.2\% of those of the constraint loop; equivalent patches written differently, as in \texttt{keepCallout} below, are not counted.

\MyPara{Anchor regions.} Table~\ref{tab:regions} breaks the constraints down by anchor region.
Constraints at branch ends are the least often violated by passing tests (20.9\%), and when they become Primary, 97.3\% of them are anchored at the faulty line.
Apart from returns, constraints after a variable definition are the least precise (51.7\%), because the definition of a variable that the faulty value flows into can often be patched to compensate; return constraints are rarely anchored at the fault (12.9\%), which supports the low weight of coarse regions.
The weights only order the third tier: with any order of its lines, Acc@1 would not change and Acc@5 would lie between 93.1\% and 93.7\%.

\input{tables/fse_rq3}

\MyPara{Case study: \texttt{keepCallout}.}
Listing~\ref{lst:case} shows \texttt{keepCallout}, in which the guard on Line~2 misses the case where no tooltip exists.
The failing test calls the method without a tooltip, and the program crashes on Line~5.
GPT-5.5 names Line~5 as its first candidate and Line~2 as its second.
Among its constraints is a branch constraint on the guard of Line~2, checked on the path that continues past it, stating \codefontsmall{self.\_tooltip is not None}; only the failing test violates it.
The constraint loop patches Line~2, guided by the branch constraint, to \codefontsmall{if not self.chart() or self.\_tooltip is None:}, which is equivalent to the developer's fix, and all tests pass.
The hypothesis check then patches Line~5 to \codefontsmall{self.\_tooltip = self.\_tooltip or self.\_add\_Callout()}, which breaks four passing tests, so the check does not confirm the first candidate.
\tool ranks Line~2 first as verified, while \toolnocf keeps the symptom first; CF makes this correction in six runs of three models.

\rqanswer{2}{Almost all constraints (99.4\%) become runtime checks, and half of those violated by failing tests state a wrong property.
Still, 90.0\% of Primary patches change the faulty line, and the hypothesis check verifies correct first answers 3.9 times as often as wrong ones.}

\subsection{RQ3: LLM Constraints and Dynamic Invariants}\label{sec:rq3}

\MyPara{Setup.}
Dynamic invariant detectors infer semantic properties without an LLM.
To compare the two sources of properties fairly, we feed both through the same pipeline.
For Daikon~\cite{ernstDaikon2001,daikon}, we built a front end that records program points at a subset of \tool's anchors, namely the function entry, every SSA definition, and the return, with parameters, locals, the return value, and one level of object fields as variables.
Daikon infers invariants from the passing tests only.
We translate them to Python expressions, drop invariants that refer to earlier states, and ground, instrument, and score them exactly like LLM constraints.
We do the same for the entry and return invariants of The Debugging Book annotator~\cite{debuggingbook}.
On the LLM side, we use the constraints of the RQ1 runs.
No row uses candidate lines or CF, which the invariant tools lack and which would bring in the LLM.

\MyPara{Results.}
Daikon infers 78.0 invariants per program (Table~\ref{tab:rq3}), 23 times as many as the 3.4 constraints of the LLMs, but reaches only 3.8\% Acc@1 and 10.0\% Acc@5.
In 451 of the 809 programs, no failing test violates any invariant, and even in the other 358 programs the faulty line is among the first five in only 22.6\% of the cases.
Where coverage gives little signal, the gap is larger: on the 185 programs with coverage collapse, LLM constraints rank the faulty line first in 40.9\%, Daikon invariants in 7.0\%, and Ochiai in 2.2\%.
The LLM sees the failing tests and states what the code should compute, while Daikon can only generalize what the passing runs did.
Adding Daikon invariants to the LLM constraints lowers accuracy for both models we tried, from 38.4\% to 31.1\% Acc@1 for gpt-oss-120b and from 24.6\% to 22.6\% for GPT-5.5.
The extra invariants add violations away from the fault.

\rqanswer{3}{Under the same pipeline, LLM constraints rank the faulty line first in 32.2\% of the programs (40.9\% under coverage collapse) and Daikon invariants in 3.8\% (7.0\%).}

\subsection{RQ4: Cost}\label{sec:rq4}

\MyPara{Setup.}
\tool logs the tokens and the wall-clock time of every LLM call and the time of every test run.
For \tool, we sum the inference call, the patch requests, and the test runs; for the two LLM baselines, the single call.
We report means per program over the runs of a model.

\MyPara{Results.}
\tool makes between 2.13 and 2.31 LLM calls per program (Table~\ref{tab:rq4}): one inference call and, on average, 1.13 to 1.31 patch requests; no program has more than five constraints, so the limit of the constraint loop never applies.
It uses 2.2 to 3.0$\times$ the tokens of \toolnoidx, and it runs the tests 3.3 to 3.7 times per program, which takes 2.8 to 3.0 seconds.
Its LLM time is 0.9 to 2.8$\times$ that of the single call.
No run had a per-call price, so we report tokens and time.

\rqanswer{4}{\tool uses 2 to 3$\times$ the tokens of one LLM question and about 3 seconds of test execution per program.}

\input{tables/fse_rq4}

%% file: tables/fse_funnel.tex
\begin{table}[t]
\centering
\begin{minipage}[t]{0.57\textwidth}
\centering
\caption{Construction of \sembugs. Each row gives the fixes that remain after the step.}
\label{tab:funnel}
\footnotesize
\begin{tabular}{@{}lr@{}}
\toprule
Step & Fixes \\
\midrule
Bug fixes in CTSSB-1M & 966{,}985 \\
in the 13 expression-level patterns & 276{,}003 \\
inside a function, in a non-test source file & 189{,}024 \\
unambiguous faulty location & 184{,}288 \\
in a non-fork repository with at least 100 stars & 18{,}253 \\
sampled, at most 300 per pattern and 2 per project & 3{,}322 \\
\midrule
same statement structure, fault inside an expression & 2{,}355 \\
kept by the LLM judge & 1{,}373 \\
\midrule
harness generated (stubs and tests) & 1{,}344 \\
verified by execution & 1{,}051 \\
after removing near-duplicates & 1{,}044 \\
no failing test crashes on the faulty line & 809 \\
\bottomrule
\end{tabular}
\end{minipage}\hfill
\begin{minipage}[t]{0.40\textwidth}
\centering
\caption{Evaluation datasets. Lines, tests, and failing tests are means per program.}
\label{tab:datasets}
\footnotesize
\begin{tabular}{@{}lrr@{}}
\toprule
 & \sembugs & QuixBugs \\
\midrule
Programs & 809 & 33 \\
Repositories & 750 & -- \\
Bug patterns & 13 & -- \\
Lines & 35.5 & 12.2 \\
Tests & 19.5 & 7.3 \\
Failing tests & 5.1 & 4.6 \\
Coverage collapse & 185 & 19 \\
\bottomrule
\end{tabular}
\end{minipage}
\end{table}

%% file: tables/fse_rq1.tex
\begin{table}[t]
\centering
\caption{Fault localization accuracy (\%). Mean over the runs of each model; \tool rows shaded; best LLM-based value per block in bold.}
\label{tab:rq1}
\fontsize{7.6pt}{8.2pt}\selectfont
\setlength{\tabcolsep}{2.3pt}
\begin{tabularx}{\textwidth}{@{}l*{10}{>{\raggedleft\arraybackslash}X}@{}}
\toprule
 & \multicolumn{4}{c}{\sembugs} & \multicolumn{3}{c}{Coverage collapse} & \multicolumn{3}{c}{QuixBugs} \\
\cmidrule(lr){2-5}\cmidrule(lr){6-8}\cmidrule(lr){9-11}
Method & Acc@1 & Acc@3 & Acc@5 & \makecell[r]{\%\\Flagged} & Acc@1 & Acc@3 & Acc@5 & Acc@1 & Acc@3 & Acc@5 \\
\midrule
\multicolumn{11}{l}{\emph{All models}} \\
\rowcolor{semlocrow} \tool & 81.4 & 91.5 & 93.4 & 35.5 & 86.3 & 94.8 & 96.5 & \textbf{74.2} & \textbf{92.4} & 96.0 \\
\toolnocf & 80.8 & 91.4 & 93.4 & 35.5 & 86.4 & 94.8 & 96.4 & 73.0 & \textbf{92.4} & 96.0 \\
\toolcand & 80.8 & 91.3 & 92.7 & 25.7 & 86.4 & 94.8 & 95.7 & 73.0 & \textbf{92.4} & 95.0 \\
\toolnoidx & \textbf{82.0} & \textbf{93.1} & \textbf{96.0} & 35.2 & \textbf{87.2} & \textbf{95.9} & \textbf{97.7} & 72.2 & 89.9 & \textbf{98.7} \\
\directllm & 50.0 & 55.5 & 55.7 & 9.8 & 59.3 & 63.1 & 63.5 & 71.0 & 76.5 & 76.8 \\
\midrule
\multicolumn{11}{l}{\emph{GPT-5.5}} \\
\rowcolor{semlocrow} \tool & 92.1 & \textbf{97.6} & 98.3 & 41.2 & 93.0 & 97.8 & 98.4 & \textbf{85.9} & \textbf{100.0} & \textbf{100.0} \\
\toolnocf & 92.0 & 97.4 & 98.2 & 41.2 & 92.6 & 97.8 & 98.2 & 85.8 & \textbf{100.0} & \textbf{100.0} \\
\toolcand & 92.0 & 97.4 & 98.1 & 30.1 & 92.6 & 97.8 & 98.2 & 85.8 & \textbf{100.0} & \textbf{100.0} \\
\toolnoidx & \textbf{92.6} & 97.4 & \textbf{98.7} & 35.5 & \textbf{93.7} & \textbf{98.6} & \textbf{99.1} & 84.8 & 95.0 & \textbf{100.0} \\
\directllm & 64.5 & 67.1 & 67.3 & 8.5 & 71.9 & 74.4 & 74.8 & 85.8 & 85.8 & 85.8 \\
\midrule
\multicolumn{11}{l}{\emph{gpt-oss-120b}} \\
\rowcolor{semlocrow} \tool & 73.9 & 86.2 & 88.8 & 34.7 & \textbf{81.1} & 92.1 & 94.6 & 66.7 & \textbf{90.9} & 94.9 \\
\toolnocf & 73.1 & 86.0 & 88.8 & 34.7 & \textbf{81.1} & 92.1 & 94.6 & 62.6 & \textbf{90.9} & 94.9 \\
\toolcand & 73.1 & 86.0 & 87.4 & 25.6 & \textbf{81.1} & 92.1 & 92.8 & 62.6 & \textbf{90.9} & 92.9 \\
\toolnoidx & \textbf{75.1} & \textbf{90.6} & \textbf{94.7} & 35.0 & 80.3 & \textbf{94.2} & \textbf{96.6} & 66.7 & 87.9 & \textbf{100.0} \\
\directllm & 46.4 & 56.0 & 56.4 & 11.2 & 55.2 & 62.2 & 62.9 & \textbf{68.7} & 83.8 & 84.8 \\
\midrule
\multicolumn{11}{l}{\emph{Qwen3.8-27B}} \\
\rowcolor{semlocrow} \tool & 76.9 & 90.4 & 92.7 & 32.3 & 82.0 & \textbf{94.4} & 96.8 & \textbf{65.7} & \textbf{88.9} & 93.9 \\
\toolnocf & 76.2 & 90.3 & 92.7 & 32.3 & 82.0 & \textbf{94.4} & 96.8 & 64.6 & \textbf{88.9} & 93.9 \\
\toolcand & 76.2 & 90.1 & 91.7 & 22.0 & 82.0 & \textbf{94.4} & 96.0 & 64.6 & \textbf{88.9} & 91.9 \\
\toolnoidx & \textbf{78.5} & \textbf{91.0} & \textbf{94.9} & 35.1 & \textbf{86.7} & 94.2 & \textbf{97.3} & 62.6 & \textbf{88.9} & \textbf{99.0} \\
\directllm & 44.4 & 47.1 & 47.2 & 9.0 & 54.4 & 56.4 & 56.8 & 63.6 & 65.7 & 65.7 \\
\midrule
\multicolumn{11}{l}{\emph{Qwen3.8-Flash}} \\
\rowcolor{semlocrow} \tool & \textbf{82.5} & 91.7 & 93.9 & 33.8 & 89.2 & 95.0 & 96.2 & \textbf{78.8} & \textbf{89.9} & 95.0 \\
\toolnocf & 81.9 & 91.7 & 93.9 & 33.8 & \textbf{89.7} & 95.0 & 96.2 & \textbf{78.8} & \textbf{89.9} & 95.0 \\
\toolcand & 81.9 & 91.7 & 93.7 & 25.3 & \textbf{89.7} & 95.0 & 95.7 & \textbf{78.8} & \textbf{89.9} & 95.0 \\
\toolnoidx & 81.6 & \textbf{93.2} & \textbf{95.6} & 35.0 & 88.1 & \textbf{96.7} & \textbf{98.0} & 74.8 & 87.9 & \textbf{96.0} \\
\directllm & 44.8 & 51.7 & 51.8 & 10.3 & 55.7 & 59.6 & 59.6 & 65.7 & 70.7 & 70.7 \\
\midrule
\multicolumn{11}{l}{\emph{Non-LLM baselines}} \\
SBFL (Ochiai) & 15.7 & 43.1 & 62.7 & 63.1 & 2.2 & 18.9 & 44.9 & 24.2 & 45.5 & 54.5 \\
SBFL (Tarantula) & 12.5 & 38.6 & 58.7 & 63.1 & 2.2 & 17.8 & 42.7 & 15.2 & 42.4 & 54.5 \\
SBFL (D*) & 15.9 & 43.5 & 62.7 & 63.1 & 2.2 & 18.9 & 44.9 & 24.2 & 45.5 & 54.5 \\
SBFL (Op2) & 16.2 & 43.9 & 62.9 & 63.1 & 2.2 & 19.5 & 44.9 & 24.2 & 45.5 & 54.5 \\
SBFL (Barinel) & 12.5 & 38.6 & 58.7 & 63.1 & 2.2 & 17.8 & 42.7 & 15.2 & 42.4 & 54.5 \\
SBFL (Jaccard) & 15.5 & 42.8 & 62.7 & 63.1 & 2.2 & 18.9 & 44.9 & 24.2 & 45.5 & 54.5 \\
Delta debugging & 1.9 & 6.7 & 12.0 & 49.0 & 6.5 & 16.8 & 28.1 & 0.0 & 3.0 & 15.2 \\
\bottomrule
\end{tabularx}
\end{table}

%% file: tables/fse_patterns.tex
\begin{table}[t]
\centering
\caption{Accuracy by bug pattern on \sembugs (\%). Mean over all runs; best value per row in bold.}
\label{tab:patterns}
\scriptsize
\setlength{\tabcolsep}{1.8pt}
\begin{tabular}{@{}lr rrr R{0.41in}R{0.41in}R{0.41in}R{0.41in}R{0.41in}R{0.41in} r@{}}
\toprule
 & & \multicolumn{3}{c}{LLM-based} & \multicolumn{6}{c}{SBFL} & \\
\cmidrule(lr){3-5}\cmidrule(lr){6-11}
Bug pattern & Programs & \tool & \makecell[r]{\tool w/o\\semantic\\indexing} & \makecell[r]{Direct\\LLM} & Ochiai & Tarantula & D* & Op2 & Barinel & Jaccard & \makecell[r]{Delta\\debugging} \\
\midrule
\multicolumn{12}{l}{\emph{Acc@1}} \\
Binary operator & 140 & 87.3 & \textbf{87.4} & 61.5 & 16.4 & 10.0 & 17.9 & 17.9 & 10.0 & 16.4 & 2.9 \\
Unary operator & 96 & \textbf{89.4} & 86.5 & 66.5 & 17.7 & 8.3 & 17.7 & 18.8 & 8.3 & 17.7 & 4.2 \\
Less specific if & 88 & \textbf{65.1} & 63.8 & 21.9 & 5.7 & 3.4 & 5.7 & 5.7 & 3.4 & 3.4 & 0.0 \\
More specific if & 82 & 58.1 & \textbf{65.4} & 24.5 & 1.2 & 1.2 & 1.2 & 1.2 & 1.2 & 1.2 & 0.0 \\
Identifier & 70 & 83.8 & \textbf{84.4} & 66.3 & 12.9 & 12.9 & 12.9 & 14.3 & 12.9 & 12.9 & 0.0 \\
Numeric literal & 59 & 87.0 & \textbf{88.1} & 49.5 & 20.3 & 16.9 & 20.3 & 20.3 & 16.9 & 20.3 & 0.0 \\
Keyword argument & 59 & 87.3 & \textbf{88.5} & 50.6 & 25.4 & 25.4 & 25.4 & 25.4 & 25.4 & 25.4 & 1.7 \\
Attribute & 49 & 85.7 & \textbf{86.2} & 58.8 & 28.6 & 26.5 & 28.6 & 28.6 & 26.5 & 28.6 & 4.1 \\
Boolean literal & 40 & \textbf{90.2} & 87.7 & 46.5 & 10.0 & 10.0 & 10.0 & 10.0 & 10.0 & 10.0 & 5.0 \\
Binary operand & 39 & \textbf{91.2} & 89.9 & 50.8 & 20.5 & 17.9 & 20.5 & 20.5 & 17.9 & 20.5 & 2.6 \\
Swapped arguments & 33 & 73.2 & \textbf{74.8} & 40.6 & 12.1 & 12.1 & 12.1 & 12.1 & 12.1 & 12.1 & 0.0 \\
Function name & 33 & 83.6 & \textbf{86.9} & 58.3 & 33.3 & 30.3 & 33.3 & 33.3 & 30.3 & 33.3 & 3.0 \\
Constant type & 21 & 86.5 & \textbf{88.1} & 47.2 & 19.0 & 14.3 & 19.0 & 19.0 & 14.3 & 19.0 & 0.0 \\
\midrule
\multicolumn{12}{l}{\emph{Acc@5}} \\
Binary operator & 140 & 93.8 & \textbf{96.2} & 66.3 & 62.1 & 60.7 & 62.1 & 62.9 & 60.7 & 62.1 & 10.0 \\
Unary operator & 96 & 97.7 & \textbf{98.7} & 72.2 & 65.6 & 50.0 & 65.6 & 65.6 & 50.0 & 66.7 & 16.7 \\
Less specific if & 88 & 86.5 & \textbf{92.4} & 23.9 & 54.5 & 52.3 & 54.5 & 55.7 & 52.3 & 53.4 & 0.0 \\
More specific if & 82 & 92.8 & \textbf{95.2} & 28.8 & 54.9 & 54.9 & 54.9 & 54.9 & 54.9 & 54.9 & 0.0 \\
Identifier & 70 & 94.0 & \textbf{97.3} & 75.8 & 65.7 & 61.4 & 65.7 & 65.7 & 61.4 & 65.7 & 8.6 \\
Numeric literal & 59 & 93.9 & \textbf{97.9} & 56.1 & 59.3 & 55.9 & 59.3 & 59.3 & 55.9 & 59.3 & 11.9 \\
Keyword argument & 59 & 93.9 & \textbf{96.9} & 54.9 & 78.0 & 76.3 & 78.0 & 78.0 & 76.3 & 78.0 & 30.5 \\
Attribute & 49 & 94.6 & \textbf{96.6} & 67.2 & 59.2 & 53.1 & 59.2 & 59.2 & 53.1 & 59.2 & 12.2 \\
Boolean literal & 40 & \textbf{94.8} & 93.8 & 51.7 & 57.5 & 55.0 & 57.5 & 57.5 & 55.0 & 57.5 & 17.5 \\
Binary operand & 39 & 96.2 & \textbf{98.1} & 55.8 & 61.5 & 59.0 & 61.5 & 61.5 & 59.0 & 61.5 & 23.1 \\
Swapped arguments & 33 & 89.4 & \textbf{90.4} & 50.0 & 60.6 & 57.6 & 60.6 & 60.6 & 57.6 & 60.6 & 15.2 \\
Function name & 33 & 94.7 & \textbf{96.5} & 65.9 & 75.8 & 72.7 & 75.8 & 75.8 & 72.7 & 75.8 & 24.2 \\
Constant type & 21 & 92.9 & \textbf{94.0} & 55.2 & 76.2 & 76.2 & 76.2 & 76.2 & 76.2 & 76.2 & 4.8 \\
\bottomrule
\end{tabular}
\end{table}

%% file: tables/fse_rq2.tex
\begin{table}[t]
\centering
\caption{Constraints and counterfactual verification on \sembugs. Shares are over all constraints.}
\label{tab:rq2}
\footnotesize
\setlength{\tabcolsep}{2pt}
\begin{tabular}{l r rrrrr rr}
\toprule
 & & \multicolumn{5}{c}{Constraints (\%)} & \multicolumn{2}{c}{Primary patches} \\
\cmidrule(lr){3-7}\cmidrule(lr){8-9}
Model & \makecell{Constraints\\per program} & Instrumented & Executed & \makecell{Violated by\\failing tests} & \makecell{Violated by\\passing tests} & \makecell{Hold on\\fixed program} & Count & \makecell{On faulty\\line (\%)} \\
\midrule
GPT-5.5 & 4.13 & 99.8 & 95.4 & 67.3 & 31.5 & 60.6 & 1127 & 89.5 \\
gpt-oss-120b & 3.15 & 99.5 & 93.8 & 43.8 & 33.0 & 59.4 & 576 & 86.6 \\
Qwen3.8-27B & 3.28 & 99.0 & 93.4 & 49.5 & 39.8 & 54.3 & 510 & 91.0 \\
Qwen3.8-Flash & 3.01 & 99.2 & 93.0 & 49.3 & 33.9 & 60.9 & 618 & 93.0 \\
\midrule
All & 3.39 & 99.4 & 94.0 & 53.6 & 34.4 & 58.9 & 2831 & 90.0 \\
\bottomrule
\end{tabular}
\end{table}

%% file: tables/fse_regions.tex
\begin{table}[t]
\centering
\caption{Constraints by anchor region on \sembugs, all models (\%). Primary: CF status of the constraint.}
\label{tab:regions}
\footnotesize
\setlength{\tabcolsep}{4pt}
\begin{tabular}{@{}l rrrr rr@{}}
\toprule
Region & Share & \makecell{Anchored at\\faulty line} & \makecell{Violated by\\failing tests} & \makecell{Violated by\\passing tests} & Primary & \makecell{Primary, anchored\\at faulty line} \\
\midrule
\texttt{LINE} & 29.7 & 80.4 & 60.7 & 37.1 & 20.9 & 94.1 \\
\texttt{AFTER\_DEF} & 25.0 & 28.6 & 44.9 & 30.7 & 2.6 & 51.7 \\
\texttt{BEFORE\_USE} & 22.3 & 43.7 & 58.7 & 42.0 & 3.3 & 86.9 \\
\texttt{AFTER\_BRANCH} & 11.8 & 53.6 & 46.1 & 20.9 & 6.6 & 97.3 \\
\texttt{ANY\_RETURN} & 5.5 & 12.9 & 54.4 & 37.8 & 1.0 & 33.3 \\
\texttt{LOOP\_TAIL} & 3.9 & 29.6 & 49.3 & 26.9 & 3.5 & 95.6 \\
Other & 1.7 & 43.9 & 48.9 & 41.6 & 2.6 & 100.0 \\
\bottomrule
\end{tabular}
\end{table}

%% file: code/case-keepcallout.tex
\begin{listing}[t]
\caption{A \sembugs fault in \texttt{keepCallout} (SimplePyScripts). The fix extends the condition on Line~2.}
\label{lst:case}
\begin{minted}[fontsize=\scriptsize, linenos, frame=lines]{python}
def keepCallout(self, point):
    if not self.chart():            # fix: if not self.chart() or not self._tooltip:
        return

    self._tooltip.setAnchor(point)
    self._callouts.append(self._tooltip)

    self._tooltip = self._add_Callout()
\end{minted}
\end{listing}

%% file: tables/fse_rq3.tex
\begin{table}[t]
\centering
\caption{Fault localization accuracy of LLM constraints and dynamic invariants (\%). All rows use the same grounding, runtime checks, and scoring, without candidate lines or CF.}
\label{tab:rq3}
\footnotesize
\setlength{\tabcolsep}{3.5pt}
\begin{tabular}{l r rrr r rrr}
\toprule
 & \multicolumn{4}{c}{\sembugs} & & \multicolumn{3}{c}{QuixBugs} \\
\cmidrule(lr){2-5}\cmidrule(lr){7-9}
Property source & \makecell{Properties\\per program} & Acc@1 & Acc@3 & Acc@5 & & Acc@1 & Acc@3 & Acc@5 \\
\midrule
Daikon invariants & 78.0 & 3.8 & 8.2 & 10.0 & & 0.0 & 12.1 & 12.1 \\
Debugging Book invariants & 4.5 & 2.2 & 2.5 & 2.6 & & 0.0 & 9.1 & 9.1 \\
\midrule
LLM constraints (all models) & 3.4 & 32.2 & 58.6 & 65.0 & & 25.5 & 55.1 & 72.0 \\
\midrule
LLM constraints (gpt-oss-120b, $T{=}0.8$) & 3.2 & 38.4 & 57.5 & 61.9 & & 39.4 & 60.6 & 69.7 \\
\quad + Daikon invariants & 81.1 & 31.1 & 50.7 & 58.1 & & 24.2 & 54.5 & 69.7 \\
LLM constraints (GPT-5.5, run 1) & 4.1 & 24.6 & 67.1 & 77.5 & & 9.1 & 57.6 & 81.8 \\
\quad + Daikon invariants & 82.1 & 22.6 & 60.9 & 73.7 & & 0.0 & 45.5 & 72.7 \\
\bottomrule
\end{tabular}
\end{table}

%% file: tables/fse_rq4.tex
\begin{table}[t]
\centering
\caption{Cost per program on \sembugs. Means over runs; tokens in thousands, time in seconds.}
\label{tab:rq4}
\footnotesize
\setlength{\tabcolsep}{3.5pt}
\begin{tabular}{@{}l rrrrr rrr rrr@{}}
\toprule
 & \multicolumn{5}{c}{\tool} & \multicolumn{3}{c}{\toolnoidx} & \multicolumn{3}{c}{\directllm} \\
\cmidrule(lr){2-6}\cmidrule(lr){7-9}\cmidrule(lr){10-12}
Model & Calls & In & Out & LLM time & Test time & In & Out & LLM time & In & Out & LLM time \\
\midrule
GPT-5.5 & 2.31 & 47.2 & 1.6 & 39.8 & 3.0 & 16.0 & 0.4 & 14.5 & 26.7 & 0.6 & 17.5 \\
gpt-oss-120b & 2.13 & 7.5 & 3.1 & 238.7 & 2.9 & 3.7 & 1.2 & 87.4 & 0.7 & 0.5 & 30.6 \\
Qwen3.8-27B & 2.20 & 8.1 & 0.6 & 11.2 & 3.0 & 3.8 & 0.2 & 12.3 & 0.8 & 0.2 & 6.7 \\
Qwen3.8-Flash & 2.21 & 8.2 & 0.8 & 23.6 & 2.8 & 3.8 & 0.2 & 15.8 & 0.8 & 0.2 & 14.4 \\
\bottomrule
\end{tabular}
\end{table}

%% file: sections/related_work.tex
\emph{Spectrum-based and dependence-based fault localization.}
SBFL ranks statements by how often failing and passing tests run them~\cite{sbflSurvey,fl}, with formulas such as Tarantula~\cite{tarantula}, Ochiai~\cite{OchiaiSbflCov,jaccardAbreu2007}, D*~\cite{dstar}, Op2~\cite{op2}, and Barinel~\cite{barinel}.
Slicing~\cite{programSlicing,dynamicSlicing} and delta debugging~\cite{zellerDD,hdd} follow dependences or minimize differences between runs.
They all rely on executed statements and their dependences; \tool's constraint violations separate even tests that run the same lines.

\emph{Mutation-based and causal fault localization.}
Metallaxis-FL~\cite{metallaxis} and MUSE~\cite{muse} mutate statements and observe how test outcomes change, and causal-inference approaches estimate the effect of executing a statement on failure~\cite{baahCausal}.
Like CF, they ask what happens when the code changes, but their mutation operators are generic and applied to many statements, whereas CF tries a few patches per program, each derived from a violated constraint or from the LLM's fault hypothesis.
For Java, ProFL~\cite{profl} reranks methods by the test outcomes of repair patches, UniVal~\cite{unival} learns the causal effect of values and predicates, and SmartFL~\cite{smartfl} models value correctness probabilistically.

\emph{LLM-based fault localization.}
AutoFL~\cite{autofl} lets an LLM navigate a repository with function calls and explain its answer, AgentFL~\cite{agentfl} splits localization among several agents, FlexFL~\cite{flexfl} works with open-source LLMs and different kinds of bug information, and LLMAO~\cite{llmao} trains adapters on LLM representations to rank lines without tests.
RGFL~\cite{rgfl} ranks files and code elements with LLM-written explanations, SieveFL~\cite{sievefl} and FaR-Loc~\cite{farloc} rank methods by coverage and retrieval, and ARISE~\cite{arise} gives LLM agents static data-flow slices.
These approaches output locations and, at most, natural-language explanations.
\tool instead turns the LLM's reasoning into constraints that are executed on every test and patches that are checked against the tests, so the evidence behind an answer can be inspected and verified.
For repair, TraceRepair~\cite{tracerepair} checks hypotheses against logged values, and many checks that repair agents pass do not test the bug~\cite{bsgva}.

\emph{Invariants and specifications.}
Daikon~\cite{ernstDaikon2001,daikon} infers likely invariants from executions.
Sahoo et al.~\cite{sahooInvariants} use invariants learned from passing inputs to select root-cause candidates, and Wang and Liu~\cite{wangDisparities} use invariants that differ between passing and failing tests as fault clues.
LLMs have been used to predict invariants~\cite{peiInvariants}, to turn natural-language intent into postconditions~\cite{nl2postcond}, and to generate formal specifications for verification~\cite{specgen}.
\tool also makes intent explicit, but anchors each property on the code graph, uses its violations as evidence, and tests the patches it suggests.

\emph{Bug datasets.}
ManySStuBs4J~\cite{manysstubs} and TSSB-3M~\cite{tssb} mine bug fixes from Java and Python projects but provide no executable tests; \sembugs adds runnable functions and generated tests.

%% file: sections/conclusion.tex
We presented \tool, which turns free-form LLM reasoning into executable and verifiable semantic constraints on the code graph.
On 809 real bugs, almost all constraints became runtime checks, their violations separated tests where coverage did not, and the tests confirmed two thirds of \tool's first answers with 93.6\% precision, at a first-rank accuracy not significantly different from that of an LLM given the same inputs.
The same artifacts can also serve test generation and repair.